\documentclass{pi2005}         % for LaTeX 2.09
\newcommand{\beq}{\begin{eqnarray}}
\newcommand{\eeq}{\end{eqnarray}}

\usepackage{graphicx}
\begin{document}
\title{Non perturbative series for the calculation of one loop integrals at finite temperature}

\authori{Paolo Amore}
\addressi{Facultad de Ciencias, \\
Universidad de Colima, \\
Bernal Diaz del Castillo no. 340,\\
Col. Villa San Sebastian, \\
Colima, Colima, Mexico}
\authorii{}
\addressii{}
\authoriii{}    \addressiii{}
\authoriv{}     \addressiv{}
\authorv{}      \addressv{}
\authorvi{}     \addressvi{}

\headauthor{Paolo Amore}
\headtitle{Non perturbative regularisation of one loop integrals at finite temperature}
\lastevenhead{Paolo Amore: Non perturbative regularisation of one loop integrals at finite temperature}
              

\maketitle
\begin{abstract}
The calculation of one loop integrals at finite temperature requires the
evaluation of certain series, which converge very slowly or can even be 
divergent. Here we review a new method, recently devised by  the author,
for obtaining accelerated analytical expressions for these series.
The fundamental properties of the new series are studied and an application 
to a physical example is considered. The relevance of the method to other
physical problems is also discussed.
\end{abstract}

\section{Introduction}     %\section*{Introduction}

In two recents papers~\cite{[1],[2]} by the author the problem of obtaining
fastly convergent series starting from slowly convergent ones was considered:
in \cite{[1]} new series representations for the Riemann, Hurwitz and more general
zeta functions were discovered; in \cite{[2]} some of the results of \cite{[1]} were
later applied to the calculation of one--loop integrals at finite temperature.

The method that we have devised in \cite{[1],[2]} is based on the Linear Delta Expansion (LDE), 
a non--perturbative technique which has been applied with success in the past to a large class of 
problems (see \cite{[3]} and references therein). The idea is to interpolate a given problem, 
which cannot be solved analytically, with a solvable problem, depending upon one or more 
arbitrary parameters. By then identifying a ``perturbation'' and performing an expansion in 
such term, a series depending upon the arbitrary parameters is found. Finally, by applying the 
``Principle of Minimal Sensitivity'' (PMS)\cite{[4]} the optimal values for the 
arbitrary parameters can be obtained. Notice that the expansions that are obtained following such procedure
are really non--perturbative, since they do not involve any parameter in the problem to be small.

This paper is organized as follows: in Section \ref{sec_2} we describe the method and apply it to 
obtain a new series representation for a generalized zeta function; in Section \ref{sec_3}
we consider the application of the method to the calculation of one--loop integrals at finite temperature;
finally, in Section \ref{sec_4} we draw our conclusions.

\section{The method}
\label{sec_2}

We consider the generalized zeta function defined as
\begin{equation}
\overline{\zeta}(u,s,\xi) \equiv \sum_{n=0}^\infty \frac{1}{(n^u+\xi)^s} \ ,
\label{eq_1}
\end{equation}
where $u s> 1$. Special cases of eq.~(\ref{eq_1}) are the Riemann zeta function, which
is obtained for $u=1$ and $\xi=1$ and the Hurwitz zeta function which corresponds
to $u=1$.

We will now prove the following
\subsection{Theorem}
It is possible to write $\overline{\zeta}(u,s,\xi)$ as
\begin{eqnarray}
\overline{\zeta}(u,s,\xi)  &=& \frac{1}{\xi^s} + 
\sum_{k=0}^\infty \frac{\Gamma(k+s)}{\Gamma(s) } \ \Psi_{k}(\lambda,u,s,\xi) 
\label{eq_2}
\end{eqnarray}
where
\begin{eqnarray}
\Psi_{k}(\lambda,u,s,\xi) &\equiv&  \sum_{j=0}^k \ \frac{(-\xi)^j}{j! (k-j)!} 
\frac{\lambda^{2 (k-j)} }{(1+\lambda^2)^{s+k}}  \zeta(us+uj) 
\label{eq_3}
\end{eqnarray}
and $\lambda^2>\frac{\xi-1}{2}$ ($\xi > 0$).

\subsubsection{Proof}

Our proof is based on the identity
\begin{eqnarray}
\overline{\zeta}(u,s,\xi) 
= \frac{1}{\xi^s} + \sum_{n=1}^\infty \frac{1}{n^{s u}} \  
\frac{1}{\left(1+\lambda^2\right)^s} \ \frac{1}{\left(1+ \Delta(n)\right)^s} 
\label{eq_4}
\end{eqnarray}
where
\begin{eqnarray}
\Delta(n) &\equiv& \frac{\xi/n^u-\lambda^2}{1+\lambda^2} \ .
\label{eq_5}
\end{eqnarray}

Provided that $\lambda^2 > \frac{\xi-1}{2}$ and $\xi>-1$, it is possible to fulfill the 
condition $|\Delta(n)| < 1$ and therefore apply the binomial theorem to obtain
\begin{eqnarray}
\frac{1}{(1+\Delta(n))^s} = \sum_{k=0}^\infty \frac{\Gamma(k+s)}{\Gamma(s) \ k!} \ \left[-\Delta(n)\right]^k \ .
\nonumber
\end{eqnarray}

Using this result in eq.~(\ref{eq_4}) we have
\begin{eqnarray}
\overline{\zeta}(u,s,\xi)  &=& \frac{1}{\xi^s} + \sum_{n=1}^\infty \sum_{k=0}^\infty \frac{\Gamma(k+s)}{\Gamma(s) \ k!} 
\ \sum_{j=0}^k \ \left( \begin{array}{c}
k \\
j \\
\end{array} \right) \frac{\lambda^{2 (k-j)}}{(1+\lambda^2)^{s+k}} \frac{(-\xi)^j}{n^{u (s+j)}}  \ .
\label{eq_6}
\end{eqnarray}

By performing the sum over $n$ one obtains
\begin{eqnarray}
\overline{\zeta}(u,s,\xi)  &=& \frac{1}{\xi^s} + 
\sum_{k=0}^\infty \frac{\Gamma(k+s)}{\Gamma(s) } \sum_{j=0}^k \ \frac{(-\xi)^j}{j! (k-j)!} 
\frac{\lambda^{2 (k-j)}}{(1+\lambda^2)^{s+k}}  \zeta(u (s+j)) \ ,
\nonumber
\end{eqnarray}
which completes our proof.

Having proved our fundamental result, eq.~(\ref{eq_2}), we now discuss some of the properties of the new
series: we first stress that the series that we have obtained is independent of the arbitrary parameter 
$\lambda$ although $\lambda$ appears explicitly in the expression. This happens because 
eq.~(\ref{eq_1}) is independent of $\lambda$ and it has just been proved that our new series, 
eq.~(\ref{eq_2}) converges to eq.~(\ref{eq_1}) provided that $\lambda^2 > (\xi-1)/2$ and $\xi>-1$.
In other words we can say that eq.~(\ref{eq_2}) describes a family of series, each corresponding to a 
different value of $\lambda$ and each converging to the same series, eq.~(\ref{eq_1}). 

We can now consider the ``truncated'' series
\begin{eqnarray}
\overline{\zeta}_{(N)}(\lambda,u,s,\xi)  &=& \frac{1}{\xi^s} + 
\sum_{k=0}^N \frac{\Gamma(k+s)}{\Gamma(s) } \ \Psi_{k}(\lambda,u,s,M^2) \nonumber
\end{eqnarray}
obtained by restricting the infinite sum to the first $N+1$ terms. Obviously 
$\overline{\zeta}_{(N)}(\lambda,u,s,\xi)$ depends upon $\lambda$ as a result of having
neglected an infinite number of terms. We can use this feature to our advantage and fix
$\lambda$ so that the convergence rate of the series is maximal.

The proper value of $\lambda$ is chosen using the PMS~\cite{[4]}, which amounts to find $\lambda$
fulfilling the equation
\begin{equation}
\frac{d}{d\lambda} \ \overline{\zeta}_{(N)}(s,\lambda) = 0 \ .
\end{equation}

A straightforward mathematical interpretation of this condition is that the value of $\lambda$ complying
with this equation also minimizes the difference~\cite{[5]} 
\begin{equation}
\Xi \equiv \left[\overline{\zeta}(u,s,\xi) - \overline{\zeta}_{(N)}(\lambda,u,s,\xi) \right]^2 \ .
\end{equation}

To lowest order, which corresponds to choosing $N=1$, one obtains the optimal value 
\begin{equation}
\lambda_{PMS}^{(1)} = \sqrt{\xi  \ \frac{\zeta( u (1+ s))}{ \zeta(s u)}} ,
\label{eq_pms}
\end{equation}
which  can be used as long as $\lambda_{PMS}^2>(\xi-1)/2$. Notice that, since $\lambda_{PMS}$ depends 
upon $\xi$ then $\overline{\zeta}_{(N)}(\lambda_{PMS},u,s,\xi)$ will not be a polynomial in $\xi$: on the
other hand, if we had chosen $\lambda=0$, then  $\overline{\zeta}_{(N)}(0,u,s,\xi)$ would be a polynomial
of $N^{th}$ order in $\xi$; in that case however the convergence of the series would be strictly limited to
the region $\xi<1$. For this reason we will refer to our accelerated series corresponding to $\lambda_{PMS}$
and to $\lambda=0$ as being ``nonperturbative'' and ``perturbative'' respectively~\footnote{Typically 
the labels ``perturbative'' and ``nonperturbative'' are used to refer to expansions in some coupling constant.
As we will see shortly in a physical application the parameter $\xi$ can be related to an inverse temperature
and not to a coupling constant.}.

\bfg[t]
\bc 
\includegraphics[height=7cm]{Fig_1.eps}
\ec
\vspace{-2mm}
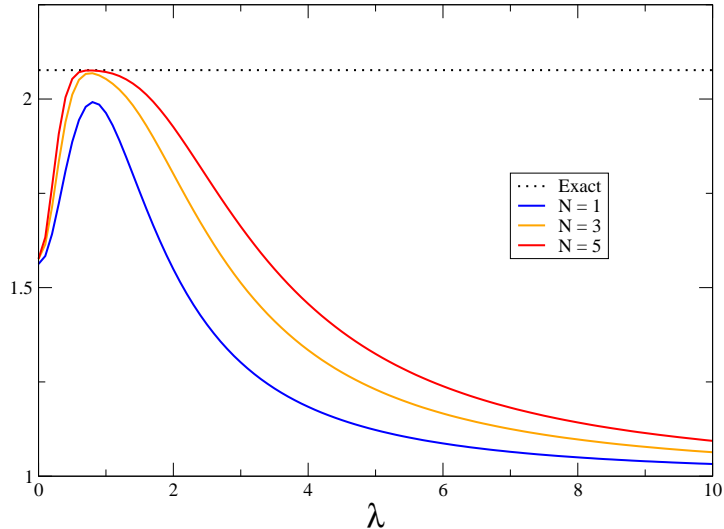
\caption{$\overline{\zeta}_{(N)}(\lambda,2,1,1)$ as  a function of $\lambda$ for $N=1,3,5$.
The dotted line corresponds to the exact value $\overline{\zeta}(2,1,1) = (1+\pi \coth \pi)/2$.
\label{fig_1}}
\efg

As an illustration we have compared in Fig.~\ref{fig_1} the exact result for
$\overline{\zeta}(2,1,1) = (1+\pi \coth \pi)/2$ (the dotted line) with the approximation 
$\overline{\zeta}_{(N)}(\lambda,2,1,1)$, with $N=1,3,5$.
The maximum of the blue curve corresponds to the value of $\lambda$ given by eq.~(\ref{eq_pms}).

\bfg[t]
\bc 
\includegraphics[height=7cm]{Fig_2.eps}
\ec
\vspace{-2mm}
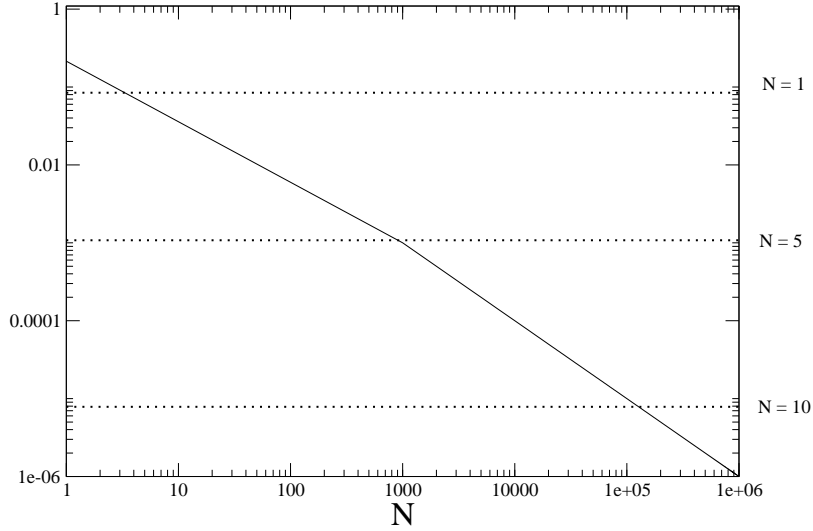
\caption{The difference $|\overline{\zeta}(2,1,1)-\sum_{n=0}^N \frac{1}{(n^2+1)}|$
as a function of $N$. The horizontal lines correspond to 
$|\overline{\zeta}(2,1,1)-\overline{\zeta}_{(N)}(\lambda_{PMS},2,1,1)|$  for $N=1,5,10$.
\label{fig_2}}
\efg     

In Fig.~\ref{fig_2} we have plotted the difference $|\overline{\zeta}(2,1,1)-\sum_{n=0}^N \frac{1}{(n^2+1)}|$
as a function of $N$ and compared them with the difference 
$|\overline{\zeta}(2,1,1)-\overline{\zeta}_{(N)}(\lambda_{PMS},2,1,1)|$  for $N=1,5,10$.
The first $10$ terms of the accelerated series, obtained by using eq.~(\ref{eq_2}) with $\lambda$
given by eq.~(\ref{eq_pms}) are able to provide the same accuracy of the first $10^5$ terms of 
the original series, eq.~(\ref{eq_1}).

%%%%%%%%%%%%%%%%%%%%%%%%%%%%%%%%%%%%%%%%%%%%%%%%%%%%%%%%%%%%%%%%%%%%%%%%%%%%%%%%%%%%%%%%%%%%%%%%%%%%%%
\section{Application: one--loop integrals at finite temperature}
\label{sec_3}
%%%%%%%%%%%%%%%%%%%%%%%%%%%%%%%%%%%%%%%%%%%%%%%%%%%%%%%%%%%%%%%%%%%%%%%%%%%%%%%%%%%%%%%%%%%%%%%%%%%%%%

One--loop integrals at finite temperature have the typical form
\begin{eqnarray}
J(m,a,b) &=& T \mu^{2 \epsilon}  \sum_{n\neq 0} \int \frac{d^Dk}{(2 \pi)^D} \ 
\frac{(k^2)^a}{\left[k^2+\omega_n^2+m^2\right]^b}  \ ,
\label{eq_3_1}
\end{eqnarray}
where $\mu$ is an energy scale brought in by dimensional regularization,
$a$ and $b$ are integers and $D \equiv 3 -2 \epsilon$ is the number of spatial dimensions.
$\omega_n = 2\pi n T$ are the Matsubara frequencies which appear in the imaginary time formalism.
We will follow the notation set in \cite{LV97} and write the expression for a general 
one--loop integral at finite temperature in the form
\beq
J(m,a,b) &=& T \mu^{2 \epsilon}  \sum_{\begin{array}{c}n=-\infty\\ n\neq 0\\ \end{array}}^\infty
\int \frac{d^Dk}{(2 \pi)^D} \ \frac{(k^2)^a}{\left[k^2+\omega_n^2+m^2\right]^b}  \nonumber 
\eeq
$\mu$ being the scale brought in by dimensional regularization, $a$ and $b$ being integers 
($a \geq 0$ and $b>0$). Following \cite{[2],LV97} we also define
\beq
K^2 \equiv \left( \frac{k}{2 \pi T}\right)^2 \  , \  \ 
M^2 \equiv \left( \frac{m}{2 \pi T}\right)^2 \   , \ \
\Omega^2 \equiv \left( \frac{\mu}{2 \pi T}\right)^2 
\nonumber
\eeq
and obtain
\beq
J(M,a,b) &=& T \ \left(2\pi T\right)^{3+2 a -2 b} \ 2 \Omega^{2 \epsilon} \ \frac{\pi^{D/2}}{(2\pi)^D} \ 
\frac{\Gamma(D/2+a)}{\Gamma(D/2)} \ \frac{\Gamma(l)}{\Gamma(b)} \ S(M,l)
\label{eq_3_2}
\eeq
where 
\beq
S(M,l) &=& \sum_{n=1}^\infty \ \frac{1}{(n^2+M^2)^l}
\label{eq_3_3}
\eeq
and $l = b-a-D/2$ and $D= 3 - 2\ \epsilon$. Depending upon the value of $l$, the series of 
eq.~(\ref{eq_3_3}) could be divergent and therefore need regularization. 

Eq.~(\ref{eq_3_3}) is clearly of the form considered in eq.~(\ref{eq_2}) and one can write 
\beq
S(M,l) &=& - \frac{1}{M^{2l}} + \zeta(2,l,M^2) \ ,
\eeq
where
\begin{eqnarray}
\overline{\zeta}(2,l,M^2)  &=& \frac{1}{M^{2l}} +  \sum_{k=0}^\infty \frac{\Gamma(k+l)}{\Gamma(l) } \  \Psi_{k}(\lambda,2,l,M^2) 
\label{eq_3_4}
\end{eqnarray}
and 
\beq
\Psi_{k}(\lambda,2,l,M^2) &\equiv& \frac{1}{(1+\lambda^2)^{s+k}} \  \sum_{j=0}^k \ \frac{(-M^2)^j}{j! (k-j)!} 
\lambda^{2 (k-j)} \zeta(2 (l+j))  \nonumber \ .
\end{eqnarray}

Once more we stress that although the series all converge to the same result independently 
of $\lambda$ (provided that $\lambda > \frac{\xi-1}{2}$), the partial sums, obtained by truncating the series 
to a finite order will necessarily display a dependence on the parameter. Such dependence, 
which is an artifact of working to a finite order, will also make  the rate of convergence 
of the different elements of the family $\lambda$-dependent. Since it is desirable to obtain the 
most precise results with the least effort, one will select the optimally convergent series 
by fixing $\lambda$ through the ``principle of minimal sensitivity'' (PMS)~\cite{Ste81}.
The solutions obtained by applying this simple criterion  display in general the highest 
convergence rate and, once plugged back in the original series, provide a 
{\sl non-polynomial} expression in the natural parameters. 

In the present case, the ``natural'' parameter in eq.~(\ref{eq_4}) is $M^2$ and to 
first--order the PMS yields
\begin{eqnarray}
\lambda_{PMS}^{(1)} &=& M \ \sqrt{\frac{\zeta( 2 (1+ l))}{ \zeta(2 l)}} ,
\label{s3_7}
\end{eqnarray}

On the other hand, if the value $\lambda = 0$ is chosen, then one obtains the series
\beq
\overline{\zeta}(2,l,M^2)  &=& \frac{1}{M^{2l}}+  \sum_{k=0}^\infty \frac{\Gamma(k+l)}{\Gamma(l) } \ 
\frac{(-M^2)^k}{k!}  \ \zeta(2 (l+k )) \ ,
\label{l0}
\end{eqnarray}
which, to a finite order yields a polynomial in $M$. Notice that such series corresponds 
to the expansion used in \cite{LV97} and converges only for  $M^2<1$: on the other hand the series 
corresponding to $\lambda_{PMS}^{(1)}$ converges for all values of $M$.

In \cite{[2]} we have studied both series, corresponding to choosing $\lambda_{PMS}^{(1)}$ and $\lambda=0$, 
showing that the first few terms of the PMS series provide a very accurate approximation to the exact 
result. For example, $\overline{\zeta}(2,3/2,1)$ calculated with the first $10^5$ terms in eq.~(\ref{eq_1}) 
gives the result $\underline{1.5124349215}0$; the same precision is reached with our PMS series 
with only $27$ terms. 

In \cite{[2]} we have used this series to evaluate the one--loop self--energy
\beq
I(m) &=& T \ \mu^{2 \epsilon} \ \sum_{n=-\infty}^{+\infty} \int \frac{d^Dk}{(2\pi)^D}  \ \frac{1}{k^2+\omega_n^2+M^2} 
\nonumber \\
&=& T \ \mu^{2 \epsilon} \  \int \frac{d^Dk}{(2\pi)^D}  \ \frac{1}{k^2+M^2} + J(M,0,1)  \ ,
\eeq
where 
\beq
J(M,0,1)  &=&  2^{-2-2\epsilon} \ \pi^{-3/2- 3 \epsilon} \ \mu^{2\epsilon} \ T^{1-4\epsilon} \ 
\Gamma(-1/2+\epsilon) \nonumber \\
&\cdot& \left[  - \frac{1}{M^{-1-2\epsilon}} + \overline{\zeta}(2,-1/2+\epsilon,M^2)\right]  \ ,
\eeq
and $d^D k = d\Omega_{D-1} \ k^{D-1} \ dk$ and  $D = 3 - 2\epsilon$. 

By using our eq.~(\ref{l0}) and retaining only the divergent terms and those independent 
of $\epsilon$ we obtain
\beq
I(m) &=& - \frac{m T}{4 \pi} + \frac{T^2}{12} - \frac{m^2}{16 \pi^2} \ \left[\frac{1}{\epsilon}  + \gamma - 
\log \frac{4\pi T^2}{\mu^2} \right] 
+ \frac{m^4 \zeta(3)}{8 (2\pi)^4 T^2} \nonumber \\
&-&   \frac{m^6\,\zeta(5)}{1024\,{\pi }^6\,T^4} + 
  \frac{5\,m^8\,\zeta(7)}{32768\,{\pi }^8\,T^6}  - 
\frac{7\,m^{10}\,\zeta(9)}{262144\,{\pi }^{10}\,T^8} + \dots
\label{pert}
\eeq
which reproduces the results of \cite{LV97}, which, however, were considered 
only up to order $m^4$. As anticipated, the formula obtained is a polynomial in $m$.
 
Using the  PMS series to {\sl first order} we obtain
\beq
J_{PMS}(M,0,1) &=&   T^2 \ \left( - \kappa_1 \ M^2   - \frac{{\zeta(3)}^{5/2} \
       {\sqrt{\zeta(3) + M^2\, \zeta(5)}}}{{\zeta(5)}^2} \right) + \kappa_2 \nonumber \\
\label{jpms0}
\eeq
where $\kappa_{1,2}$ are the constants of integration {\sl independent} of $M$:
\beq
\kappa_1 &=& \frac{1}{4} \left[ \frac{1}{\epsilon} + \gamma - \log \left(\frac{4 \pi T^2}{\mu^2}\right)-
\frac{2\,{{\zeta}(3)}^2}{{\zeta}(5)}\right] \nonumber \\
\kappa_2 &=& \frac{1}{12} + \frac{{{\zeta}(3)}^3}{{{\zeta}(5)}^2} \nonumber \ .
\eeq

As mentioned before the PMS result is non--polynomial in $M$ and at a given order contains 
more information than the corresponding result for $\lambda=0$ obtained at the same order. 
This can be verified by  Taylor expanding eq.~(\ref{jpms0}) in $M$ and comparing it with
eq.~(\ref{pert}), which is calculated to order $M^8$:
\beq
J(M,0,1) &\approx&   - \frac{m T}{4 \pi} + \frac{T^2}{12} - \frac{m^2}{16 \pi^2} \ \left[\frac{1}{\epsilon}  + \gamma - 
\log \frac{4\pi T^2}{\mu^2} \right] 
+ \frac{m^4 \zeta(3)}{8 (2\pi)^4 T^2}- \frac{m^6\,\zeta(5)}{1024\,{\pi }^6\,T^4}\nonumber \\
&+&   \frac{5\,m^8\,{\zeta(5)}^2}
   {32768\,{\pi }^8\,T^6\,\zeta(3)} - 
  \frac{7\,m^{10}\,{\zeta(5)}^3}
   {262144\,{\pi }^{10}\,T^8\,{\zeta(3)}^2} + \dots
\eeq

Not only eq.~(\ref{jpms0}) reproduces correctly  the terms going as $m^4$ and $m^6$ of the perturbative series
but it also provides {\sl all} the coefficients of the higher order terms apart from factors 
depending on the Riemann $\zeta$ function evaluated at odd integer values.

These properties has been used in \cite{[2]} to obtain a variational acceleration of the series for the thermodynamic 
potential, which has then been compared with the high-temperature series of Haber and Weldon~\cite{HW82}, showing that
the PMS result to a finite order provides an accurate approximation valid also to low temperature.

\section{Conclusions}
\label{sec_4}

We have summarized some of the results recently obtained by the author and contained in \cite{[2]}: a given slowly convergent 
 can be transformed into a fastly convergent series by introducing into the series an arbitrary parameter
and by then carefully identifying a term in which to expand. This procedure in general provides series which converge very
fast, typically geometrically, and for all values of the parameters. In the present case the natural parameter
in the problem is the inverse temperature: we have compared our results with similar results in the literature, amounting 
to a high temperature expansion, showing that the present approach provides quite precise result at high and low temperature.
Possible further applications of the ideas exposed in this paper and in \cite{[2]} include the calculation of the Casimir effect, which
is presently being carried out~\cite{AAH}.

\bigskip
{\small The author acknowledges support of CONACYT grant 40633.}

\end{document}